\documentclass[useAMS,usenatbib,psfig]{mn2e}

\usepackage{psfig}
\usepackage{times}

\newcommand{\bd}{\begin{displaymath}}
\newcommand{\ed}{\end{displaymath}}
\newcommand{\be}{\begin{equation}}
\newcommand{\ee}{\end{equation}}

\title{No evidence for a different accretion mode for all 3CR FR I radio galaxies}

\author[X. Cao \& S. Rawlings]
{ Xinwu Cao$^{1,2}$
\thanks{E-mail: cxw@center.shao.ac.cn},
Steve Rawlings$^2$\\
$^1$ Shanghai Astronomical Observatory,
Chinese Academy of Sciences, 80 Nandan Road, Shanghai, 200030, China\\
$^2$ Department of Astrophysics, University of Oxford, Oxford, OX1
3RH}

\date{Accepted 2004 January 9. Received 2003 December 17; in original form 2003 September 30}

\pagerange{\pageref{firstpage}--\pageref{lastpage}}
\pubyear{}

\begin{document}

\maketitle
\label{firstpage}

\begin{abstract}

We have analyzed the optical and radio properties of a sample of
3CR FR I radio galaxies which have Hubble Space Telescope imaging
capable of detecting optical cores. The jet powers of the FR I
radio galaxies are estimated from their low-frequency radio
luminosities, and the optical core luminosity is taken as an upper
limit on the emission from any unobscured accretion disc. We argue
that if the accretion discs in these sources are assumed to be
advection dominated accretion flows (ADAFs), or adiabatic
inflow-outflow solution (ADIOS) flows, then the Blandford-Znajek
mechanism \citep{bz77} provides insufficient power to explain the
high radio luminosities of at least a third, and perhaps all, of
the sample. We suggest instead that a significant fraction { (the
'high-jet-power' third)}, and perhaps { most}, of the 3CR FR I
radio galaxies have normal accretion discs, but that their optical
cores can be hidden, with any HST-detected optical synchrotron
emission coming from jets on scales larger than the obscuring
material. A normal accretion disc hypothesis, at least for the
high-jet-power third of the 3CR FR Is, explains why narrow-line
luminosity correlates with radio luminosity. It also explains why
one object in the sample (3C 386) has an observed broad-line
nucleus. We conclude that there is no evidence to suggest that
there is a difference in accretion mode between FR I and FR II
radio galaxies.

\end{abstract}


\section{Introduction}

FR I radio galaxies (defined by edge-darkened radio structure)
have lower radio power than FR II galaxies (defined by edge-brightened
radio structure due to compact jet-terminating hot spots) \citep{fr74}.
What causes the difference between FR I and FR II radio galaxies is still
unclear. The different explanations of the
division of FR I and FR II radio galaxies invoke either the interaction
of the jet with the ambient medium or the intrinsic nuclei
properties of accretion and jet formation processes (e.g.,
Bicknell 1995; Gopal-Krishna \& Wiita 1988; Reynolds et al. 1996;
Baum, Zirbel \& O'Dea 1995; Blundell \& Rawlings 2001). In the
frame of unification schemes of active galactic nuclei (AGNs), FR
I radio galaxies are believed to be the misaligned BL Lac objects,
and FR II radio galaxies correspond to misaligned flat-spectrum
radio quasars (see Urry \& Padovani 1995 for a review). The
central structure and activity of radio galaxies are important for
understanding the physics at work in their nuclei.

Recently, the optical nuclei of low-redshift radio galaxies have
been observed by the Hubble Space Telescope (HST), and central
compact optical cores detected in many cases
\citep{cc99,c99,c02,v02}. The fact that the optical fluxes of the
central compact cores in 3CR FR I radio galaxies correlate
linearly with their radio core fluxes, and the lack of scatter in
this correlation, argues for a synchrotron origin for the optical
emission of the cores in FR I radio  galaxies \citep{c99}. FR II
galaxies, typically, show brighter optical cores than FR I
galaxies and a more complex correlation between radio and optical
core luminosities suggests that the disc emission may contribute a
significant fraction of their optical core fluxes \citep{c02}.
\citet{c99} estimated the upper limits on the disc emission of FR
I galaxies from the observed optical core fluxes and found that
the optical core luminosities of FR I radio galaxies are far lower
than the Eddington case ($L_{\rm Edd}$), if the black holes in FR
I radio galaxies have masses around 10$^9$~${\rm M_\odot}$;
typically $\lambda L_{\rm c,opt}/L_{\rm Edd}\le 10^{-4}$, $\lambda
L_{\rm c,opt}/L_{\rm Edd} \simeq 10^{-6}$ in the case of the
archetypal nearby FR I galaxy M87.

In the frame of unification schemes, the viewing angle of FR I
galaxies is, on average, large, and it is therefore to be expected that any
geometrically and optically thick torus present could hide the
nucleus and broad-line emission in most cases \citep{up95}. In the receding
torus model \citep{l91}, in particular, the optical emission from a
comparatively weak disc is likely to be hidden at all but very small angles
to the line of sight, as the torus is predicted to lie extremely close to
the nucleus \citep{h96}. However, \citet{c99} argued that
high detection rate of optical central compact cores of FR I radio
galaxies ($\sim$85\%) in the 3CR catalogue suggests such a torus is not
present in FR I galaxies. For example, the linear resolution achieved
by their HST observations at the distance of M87 corresponds to $\sim$6 pc,
implying that, in this source at least, the synchrotron radio and optical
core emission is not being hidden by a torus, so if one is present it must
be less than 6 pc in size. Constraints on the size of any obscuring torus
clearly worsen, however, for the
more distant, powerful FR I radio galaxies in the
3CR sample.

The division between FR I and FR II radio galaxies is clearly
shown by a line in the plane of the optical luminosity of the host
galaxy and the total radio luminosity by \citet{lo96}. The FR I/II division line is a function of the
optical luminosity of the host galaxy (roughly proportional to the optical
luminosity). The total radio luminosity of almost all FR II radio
galaxies lies above the division line, while  FR I radio galaxies
are in the region below the line.  They suggested that the
accretion rate may play an important role in the division between
FR I and FR II galaxies. Recently, \citet{gc01} used the optical luminosity of the host galaxy to
estimate the mass of its central black hole, and used the radio luminosity
to derive the jet power.  They proposed that the FR I-FR II separation can be
interpreted as occurring at a constant ratio
between the jet power and the black hole mass. This suggests that the
FRI/FRII division is linked to physics on very small scales, and potentially
suggests a difference in accretion mode.

For radio galaxies and quasars, their optical/UV continuum is a
mixture of the emission from the jet and from the disc, which prevents
estimating the emission from the disc directly from
optical/UV continuum. Moreover, obscuration of the accretion disc is
known to happen in some radio galaxies \citep{sww2000}
so an orientation-independent
estimator of the nuclear luminosity is needed.

A widely adopted approach to estimate the accretion
disc emission in radio galaxies and quasars is to use their narrow
emission lines (e.g., Rawlings \& Saunders 1991; Xu, Livio \& Baum
1999), or broad lines (e.g., Celotti, Padovani \& Ghisellini 1997;
Cao \& Jiang 1999; 2001), which are ionized by the photons emitted
from the disc (although photo-ionization by radiative shocks
driven by the radio source can also play an important role: e.g.,
Inskip et al. 2002). The applicability of this approach to
FR I radio galaxies has been questioned \citep{zb95} but it is at least
roughly in accord with the correlations between line and optical luminosities
seen in high-spatial-resolution observations of nearby FR I radio galaxies
(Verdoes Kleijn et al.\ 2002).

The cosmological parameters
$H_0=50~ {\rm km~s^{-1}~Mpc^{-1}}$ and $q_0=0.5$ have been adopted
in this paper. This is not now the conventional choice, but as we are
studying only low-redshift objects, this has no significant bearing on
the results of this paper.

\section{Sample}

\citet{c99} processed the data of HST
observations in the public archive for 32 of all 33 FR I radio
galaxies belonging to the 3CR catalogue (3C76.1 has not been
observed). In this work, we use these 33 FR I radio galaxies
belonging to the 3CR catalogue as our sample. Key data on these
sources are listed in Table 1. \citet{c02} detected central
compact cores unresolved at the HST resolution in 22 radio FR I
galaxies. Five sources show larger optical cores ($>0.15''$),
which are different from those sources with detected central
compact cores ($0.05''-0.08''$). For these five sources, only
upper limits on the central compact core emission are given. The
remaining five sources have complex nuclear optical structures. In
four of these five sources, bright compact knots are resolved and
offset from the center of the galaxy. The optical image of 3C315
has an elongated structure and no central source is seen. No
reliable optical core flux is available for these five sources.

\section{Central black hole mass}

In order to estimate the central black hole masses of FR I
galaxies, we use the relation between host galaxy absolute
magnitude $M_R$ at $R$-band and black hole mass $M_{\rm bh}$
proposed by \citet{md02}
 \be
 \log_{10}(M_{\rm bh}/{\rm M_{\odot}})=-0.50(\pm0.02)M_R -2.96(\pm0.48).
 \label{mrmbh}
\ee The derived black hole masses of all sources in the sample are
listed in Table 1. For M87(3C274), its central black hole mass has
been estimated to be $\simeq 3\times 10^9 {\rm M_\odot}$
\citep{f94,h94,m97}.

\section{Jet power and its origin}

The relation  between jet power and radio luminosity proposed by \citet{w99} is
\be Q_{\rm jet}\simeq 3\times
10^{38}f^{3/2}L_{\rm 151}^{6/7}~ {\rm W}, \label{qjetrad}
\ee
where $L_{\rm 151}$ is the total radio luminosity at 151 MHz in units of
10$^{28}$ W~Hz$^{-1}$~sr$^{-1}$.
\citet{w99} have argued that the normalization
is very uncertain and
introduced the factor $f$ to account for these uncertainties. They
use a wide variety of arguments to suggest that  $1\leq f\leq 20$.
\citet{br00} argued that $f\sim 10$ is a likely
consequence of the evolution
of magnetic field strengths as radio sources evolve. This would mean
that the bulk kinetic (jet) and radiative (quasar) outputs of radio-loud
AGN are of similar magnitude (Fig. 7 of Willott et al. 1999). This
relation was proposed for radio FR II galaxies and quasars. We
adopt this relation to estimate the power of jets in FR I radio
galaxies which is an approximation (because the efficiency factor $f$
for FR Is may well be systematically different), but one that is probably
good to an order of magnitude. For example, we use this
relation to estimate the power of the jet in M87, we find $Q_{\rm
jet}\sim 6\times 10^{35}$~W in the case of $f=1$, which, for reasonable
$f$  is consistent with
previous estimates (e.g., \citet{bb99a}
found $10^{36-37}$~W; \citet{y02}
derived the temperature and pressure of the gas in its inner radio
lobes ($\leq~1.5$~kpc) from the Chandra X-ray observation. The
cavity of the X-ray gas is assumed to be created by the jet and
its age is estimated from the free-fall timescale. The minimum
power of the jet in M87 is therefore estimated as $\sim 3\times
10^{35}$~W).

It is evident that the accretion rates of FR I galaxies are very
low. It has been suggested, therefore, that their accretion discs
are probably in an  advection dominated accretion flow (ADAF)
state \citep{gc01}. An ADAF radiates inefficiently and is very
hot. Most gravitational energy of the accreting matter released is
carried into the black hole. Unlike the optically thick accretion
disc case, there is a theoretical upper limit on the accretion
rate for an ADAF \citep{ny95}. The optically thin ADAF will not
exist and it will transit to an optically thick disc, if its
accretion rate is greater than the critical one $\dot m_{\rm
crit}=\dot M_{\rm crit}/\dot M_{\rm Edd}$ (see Narayan 2002 for a
review). The exact value of the critical accretion rate $\dot
m_{\rm crit}$ is still unclear, depending on the value of the disc
viscosity parameter $\alpha$, {i.e., $\dot{m}_{\rm crit}\simeq
0.28\alpha^2$\citep{mah97}.} \citet{gc01} estimated the power of
the jets in radio galaxies from their radio luminosity. They
proposed that the division between FR I and FR II radio galaxies
can be expressed as a constant accretion rate between $\sim
10^{-2}$-$10^{-3}$ of the Eddington one on the assumption of a
constant conversion of the jet power to bolometric luminosity.
This is consistent with the critical accretion rate expected by
ADAF models in the sense that FR I radio galaxies could be radio
sources whose central engines are in the ADAF state, and FR IIs
could be radio sources whose central engines are standard,
optically-thick accretion discs. It is already known that there
would have to be exceptions to any such rule: 3C386, an FR I in
our sample (Table 1), has a relatively bright compact optical core
and a broad-line nucleus (Simpson et al.\ 1996); there is at least
one example of an optically bright quasar with FRI-like radio
structure (Blundell \& Rawlings 2001).

If the black holes in FR I galaxies are spinning rapidly, the
rotational energy of the black holes is expected to be transferred
to the jets by the magnetic fields threading the holes, namely,
the Blandford-Znajek mechanism \citep{bz77}. The power extracted
from a maximal spinning black hole was calculated by \citet{an99}. Here, we calculate the jet power in a
similar way. The power extracted from a spinning black hole is
given by (e.g. Ghosh \& Abramowicz 1997; Macdonald \& Thorne 1982)
 \be
L_{\rm BZ} = {1 \over 32} \omega_{\rm F}^2 B_\bot^2 R_{\rm h}^2 c
 a^2, \label{lbz}
\ee
for a black hole of mass $M_{\rm bh}$ and dimensionless
angular momentum $a$, with a magnetic field $B_\bot$ normal to the
horizon at $R_{\rm h}$.  $R_{\rm h}$ is proportional to the black
hole mass $M_{\rm bh}$ and is a function of $a$. Here the factor
$\omega_{\rm F}^2 \equiv \Omega_{\rm F} (\Omega_{\rm h} -
\Omega_{\rm F}) / \Omega_{\rm h}^2$ depends on the angular
velocity of field lines $\Omega_{\rm F}$ relative to that of the
hole, $\Omega_{\rm h}$. In order to estimate the maximal power
extracted from a spinning black hole, we adopt $\omega_{\rm F} =
1/2$.

The magnetic field threading the black hole is maintained by the
currents in the accretion disc surrounding the hole
\citep{ga97,l99}. The pressure of an ADAF is given by \citep{ny95}
\be p=1.71\times 10^{7}\alpha^{-1}c_1^{-1}c_3^{1/2}m^{-1}\dot m
~r^{-5/2} {\rm N~m^{-2}}, \label{pressure} \ee where the
dimensionless quantities are defined as follows: \bd m={\frac
{M_{\rm bh}} {\rm M_{\odot}}},~~~r={\frac {Rc^2}{2GM_{\rm
bh}}},~~~\dot m={\frac {\dot M}{\dot {M}_{\rm Edd}}}, \ed and \be
\dot{M}_{\rm Edd}={\frac {L_{\rm Edd}}{\eta_{\rm
eff}c^2}}=1.39\times 10^{15} m~~{\rm kg~s^{-1}}. \ee Here the
standard value of the accretion efficiency factor $\eta_{\rm
eff}=0.1$ is adopted. The parameters $c_1$ and $c_3$ are
\citep{ny95} \be c_1={\frac
{(5+2\epsilon^{\prime})}{3\alpha^2}}g(\alpha, ~\epsilon^{\prime})
\ee and \be c_3={\frac
{2(5+2\epsilon^{\prime})}{9\alpha^2}}g(\alpha,
~\epsilon^{\prime}), \ee where \be \epsilon^{\prime}={\frac
{1}{f_{\rm adv}}}\left( {\frac {5/3-\gamma}{\gamma-1}}\right) \ee
and \be g(\alpha,~\epsilon^{\prime})=\left[1+{\frac
{18\alpha^2}{(5+2\epsilon^{\prime})^2}} \right]^{1/2}-1. \ee The
parameter $f_{\rm adv}$, which lies in the range 0$-$1, is the
fraction of viscously dissipated energy which is advected;
$\gamma$ is the ratio of specific heats. So, the value of the
parameter $\epsilon^\prime$ is in the range of $0-1$. As done by
\citet{an99}, we assume $B_\bot\simeq~B$, $p_{\rm
mag}=B^2/8\pi\sim~\alpha p$, and two parameters $\alpha=1$ and
$\epsilon^\prime=1$ are adopted to maximize the pressure (see Eq.
\ref{pressure}). It is obvious that the magnetic pressure is
proportional to the dimensionless accretion rate $\dot m$ (see Eq.
{\ref{pressure}). As the accretion rate of an ADAF should be less
than the critical one $\dot m_{\rm crit}$, we can calculate the
maximal jet power extracted from a spinning black hole of mass
$M_{\rm bh}$ and dimensionless angular momentum $a$ from Eq.
(\ref{lbz}), if $\dot m=\dot m_{\rm crit}$ is substituted into Eq.
(\ref{pressure}). The results of the extreme case for $a=1$ have
already been given in \citet{an99}.

The ADAF solutions may be modified to adiabatic inflow-outflow
solutions (ADIOSs), if a powerful  wind is present to carry away
mass, angular momentum and energy from the accreting gas
\citep{bb99}. In this case, the accretion rate of the disc is a
function of radius $r$ instead of a constant accretion rate along
$r$ for a pure ADAF. For ADIOSs, the gas swallowed by the black
hole is only a small fraction of the rate at which it is supplied,
as most of the gas is carried away in the wind before it reaches
the black hole. The accretion rates at the inner edge of the disc
for ADIOSs would be at least as low as that required by pure ADAF
solutions, and they have similar structure at the inner edge of
the disc \citep{ccy02}. The jet power extracted from a spinning
black hole by the Blandford-Znajek mechanism for ADIOS cases is
therefore similar to that for pure ADAF cases, if they have
similar accretion rates at the inner edge of the disc.

\begin{table*}
 \begin{flushleft}
  \caption{Data of the sources.}
  \begin{tabular}{lcrrccrccc}\hline
Source & z & $\log_{10} L_{\rm c,opt}$ & $\log_{10} L_{\rm c,5G}$ & $m_R$ &
Refs. & $\log_{10} L_{\rm line}$ & Refs. & $\log_{10} Q_{\rm jet}/f^{3/2}$
& $\log_{10}  M_{\rm bh}/{\rm M_{\odot}}$\\
(1) & (2) & (3) & (4) & (5) &(6) & (7) & (8) & (9) & (10)\\ \hline
 3C~28$^{\rm a}$ & 0.1953 & $<$30.76 & $<$22.48  & 18.34 & d96  & 35.07 & ZB95 & 37.1 & 8.1\\
 3C~29   & 0.045  &  30.71 &   23.90 & 13.02  & G00  &   33.69 & ZB95 & 36.0 & 9.1\\
 3C~31   & 0.0169 &  30.27 &   23.05 & 11.92 & S73  &    33.18 & V02 & 35.3 & 8.6\\
 3C~66B  & 0.0213 &  30.99 &   23.55 &  12.55 & M99  &   33.64 & V99 & 35.8 & 8.5\\
 3C~75   & 0.0232 &  ... &   22.95 &  11.88 & S73  &  32.64 & ZB95 & 35.7 & 9.0\\
 3C~76.1 & 0.0325 &  not observed &  22.55 & 14.06 & S73 & ... &  & 35.6 & 8.2\\
 3C~78   & 0.0287 &  31.93 &   24.53 &  12.50 & M99  &  34.02 & ZB95 & 35.7 & 8.9\\
 3C~83.1 & 0.0251 &  29.59 &   22.75 &  12.39  & S73  &  ...   &      & 35.8 & 8.8\\
 3C~84   & 0.0176 &  32.31 &   25.75 &  11.03 &  G77 &   35.21 & ZB95 & 35.9 & 9.1\\
 3C~89$^{\rm a}$   & 0.1386 &  $<$30.25 &   24.58 & 16.55 & d96  &  34.68 & ZB95 & 37.0 & 8.6\\
 3C~264   & 0.0217 &  31.37 &   23.60 & 13.03  & M99  &   33.39 & ZB95 & 35.6 & 8.3\\
 3C~270   & 0.0075 &  29.09 &   22.87 & 10.16  & S73  &   $<$32.46 & ZB95 & 35.1 & 8.6\\
 3C~272.1 & 0.00334 & 29.45 &   21.94 & 9.26  & S73  &   32.67 & ZB95 & 34.1 & 8.2\\
 3C~274   & 0.0044 &  30.51 &   23.51 & 9.03  & S73  &   33.77 & ZB95 & 35.8 & 9.5\\
 3C~277.3$^{\rm a}$ & 0.0853 &  30.69 &   23.58 & 15.71 & E85  &   34.58 & JR97 & 36.3 & 8.5\\
 3C~288$^{\rm a}$   & 0.246  &  31.31 &   24.84 & 17.32  & d96  &   ...  &    & 37.4 & 8.9\\
 3C~293   & 0.045 &  ... &   23.93  &  13.68  & S73  &  34.84   & ZB95   & 36.0 & 8.9\\
 3C~296   & 0.0237 &  29.92 &   23.27 &  12.18  & M99  &  ...   &    & 35.5 & 8.8\\
 3C~305   & 0.0411 &  ... &   23.32 &  13.68  & M99  &  33.41   & ZB95   & 36.0 & 8.7\\
 3C~310$^{\rm a}$   & 0.054  &  30.65 &   23.99 & 15.66  & M99  &   ...   &    & 36.6 & 8.0\\
 3C~314.1$^{\rm a}$ & 0.1197 &  $<$30.79 &   $<$22.77 & 17.91  & d96  &   ...   &    & 36.6 & 7.8 \\
 3C~315$^{\rm a}$    & 0.1083 &  ... &  24.86 & 15.84$^{\rm c}$ & LL84  &    ... &  & 36.8 & 8.7  \\
 3C~317   & 0.0345 &  30.70 &   24.29 & 11.71 & G00  &    34.94 & ZB95 & 36.2 & 9.5\\
 3C~338   & 0.0304 &  30.61 &   23.62 & 12.20  & S73  &   34.08 & ZB95 & 36.2 & 9.1\\
 3C~346$^{\rm a}$   & 0.162  &  32.45 &   25.36 & 16.98  & d96  &   35.35 & ZB95 & 36.9 & 8.8\\
 3C~348$^{\rm a}$   & 0.154  &  30.94 &   23.98 & 16.13  & E85  &   33.73 & T93 & 38.1 & 8.9\\
 3C~386   & 0.017  &  32.24 &   22.24 & 12.21  & M99  &   $<$33.45 & S96 & 35.4 & 8.5\\
 3C~424$^{\rm a}$   & 0.127  &  $<$31.04 &   24.07 & 17.05$^{\rm b}$  & ZB95  &   ... &    & 36.7 & 8.3\\
 3C~433$^{\rm a}$    & 0.1016 &  ... &   23.33 & 14.82$^{\rm c}$   & L85   &  35.34 & ZB95  & 37.1 & 9.1\\
 3C~438$^{\rm a}$   & 0.29   &  $<$31.21 &   24.73 & 18.24 & d96  &   35.17 & ZB95 & 37.8 & 8.6\\
 3C~442$^{\rm a}$   & 0.0263 &  29.44 &   21.77 &  14.07 & M99 &   33.85 & ZB95 & 35.7 & 8.0\\
 3C~449   & 0.0171 &  30.36 &   22.66 & 13.05   & M99  &  32.87 & K02 & 35.2 & 8.0\\
 3C~465   & 0.0302 &  30.88 &   24.02 & 13.14  & S73  &   33.23 & M96 & 36.1 & 8.6\\
\hline
\end{tabular}
\end{flushleft}
\end{table*}
\begin{table*}
 %\centering
\begin{minipage}{170mm}
Notes for the table 1. $\rm a$: the sources for which the jet power
is greater than the maximal jet power extracted from a rapidly
spinning black hole with $a=0.95$ and $f=1$; $\rm b$: the $R$-band magnitude
is converted from the $V$-band magnitude assuming $V-R=0.61$
\citep{f95}; $\rm c$:  the $R$-band magnitude is converted from the
$K$-band magnitude assuming $R-K=2.5$ \citep{d03}; Column (1):
source name, Column (2): redshift, Column (3): $\log_{10}$ optical
core luminosity  (W~\AA$^{-1}$), Column (4): $\log_{10}$ radio core
luminosity at 5 GHz (W~Hz$^{-1}$), Column (5): rest frame $R$-band
magnitude, Column (6): references for R-band magnitude, Column
(7): $\log_{10}$ luminosity of narrow-lines H$\alpha$+[N\,{\sc
ii}] (W), Column (8): references for the line emission, Column
(9): $\log_{10}$ jet power estimated from total low-frequency
radio luminosity (W),
Column (8): the derived black hole masses \\
References: d96: \citet{d96}, E85: Eales
\citet{e85}, G77: \citet{g77}, G00: \citet{g00},
JR97: \citet{jr97}, LL84: citet{ll84}, L85: \citet{l85}, M96: \citet{m96}, M99: Martel, Baum \& Sparks
\citet{m99}, S73: \citet{s73}, S96: \citet{s96}, T93: \citet{t93}, V99:
\citet{v99}, V02: \citet{v02}, ZB95: \citet{zb95}.
 \end{minipage}
\end{table*}

\section{Results and discussion}

The derived jet powers and black hole masses for 33 FR I galaxies
are listed in Table 1. We searched the literature for flux data on
the narrow emission lines H$\alpha$+[N\,{\sc ii}].  For those
where the flux of H$\alpha$+[N\,{\sc ii}] is unavailable, we use
[O\,{\sc iii}] flux multiplying a factor 0.83 to convert to
H$\alpha$+[N\,{\sc ii}] flux as done by \citet{zb95}. We found 25
sources with narrow-line emission data.

\subsection{Jet power versus black hole mass}

As discussed in Sect. 4, the uncertainties in the jet power
estimate are included in the factor $f$ in Eq. (\ref{qjetrad}). We
plot the relation between the jet power $Q_{\rm jet}$ and black
hole $M_{\rm bh}$ in Fig. \ref{mqj} for $f=1$. As discussed in
Sect. 4, the maximal jet power extracted from a spinning black
hole with $M_{\rm bh}$ and $a$ can be calculated for an ADAF
surrounding the hole. {The three-dimensional MHD simulations
suggest the viscosity $\alpha$ in the discs to be $\sim 0.1$
\citep{a98}, or $\sim 0.05-0.2$ \citep{hb02}.} We adopt $\dot
m_{\rm crit}=0.01$ in our calculations, {corresponding to
$\alpha\simeq 0.2$, which is a conservative choice.} The
calculated maximal jet power is plotted in Fig. \ref{mqj}. We find
that the jet power for more than one-third of the sources in the
sample is greater than the maximal jet power calculated for
$a=0.95$ (13 of 33 sources, hereafter referred to as
high-jet-power sources, and the remainder as low-jet-power
sources). For some FR I sources, their jet powers can be as high
as $0.01 L_{\rm Edd}$. It indicates that if the Blandford-Znajek
mechanism is the only one in operation then it would be unable to
power the jets in these sources. In the case of $f=10$, the
relation between the jet power $Q_{\rm jet}$ and black hole
$M_{\rm bh}$ is plotted in Fig. \ref{mqj10}. The jet power $Q_{\rm
jet}$ is about 30 times that for $f=1$, which would greatly
strengthen the conclusion that the Blandford-Znajek mechanism is
unable to produce the jets in the high-jet-power sources.
Furthermore, it would suggest that the mechanism may have
difficulty powering any of the 3CR FR I sources. Anyway, hereafter
we use, conservatively, $f=1$ in the estimates of jet power
$Q_{\rm jet}$ (values for the case $f\ne 1$ can be evaluated by
multiplying by $f^{3/2}$). {The uncertainty in $f$ is
therefore much greater than any uncertainty in $\dot{m}_{\rm
crit}$ unless $\alpha\sim 1$ which seems inconsistent with the MHD
simulations \citep{a98,hb02}.}

One possible explanation may be that the jets in these radio galaxies
are not primarily accelerated by the magnetic field lines threading the
spinning black holes, for example, they may be accelerated by the field
lines of the accretion disc \citep{bp82}, or by the combination of
the two mechanisms. It has in fact been proposed that the jet is
accelerated more efficiently by the field lines threading the accretion disc
(Livio, Ogilvie \& Pringle 1999; but see also more detailed
calculations in \citet{cao02}). Even if this were the case, it implies
that almost all released gravitational energy of the accretion
matter is carried away by the jets in those sources with $Q_{\rm
jet}\sim 0.01L_{\rm Edd}$ for $f=1$. For a high value of $f$, for example,
$f=10$, the jet power can be as high as $\sim 0.3L_{\rm Edd}$ for some
sources. It is hard to see how the accretion discs in these sources
would remain in an ADAF state.
However, we cannot easily rule out the possibility that ADIOS flows are
present in these sources, if the accretion rates of the flows are close
to the Eddington value and the flows extend to
large radii \citep{bb99}. In this case, most of the mass and energy of the
accreting gas may be  carried away in the jet, and any optical
emission from the accretion flow could be very faint.

Another possibility is that the accretion discs in these
high-jet-power sources are not in ADAF state. Their accretion
rates are in fact higher than the critical accretion rate $\dot
m_{\rm crit}$. In this case, the accretion discs are expected to
be the standard optically thick ones, and their jet power can be
higher than the maximal jet power expected to be extracted by the
Blandford-Znajek mechanism for ADAF cases. We note from Figs.\ 2
\& 3 that the one object in our sample with a broad-line nucleus
(3C386) is not one of the extreme high-jet-power sources, implying
that standard accretion discs in {high-jet-power} FR I sources
may be common in our sample, although we would need a mechanism to
explain why they cannot easily be seen in the HST optical images.

\begin{figure}
\centerline{\psfig{figure=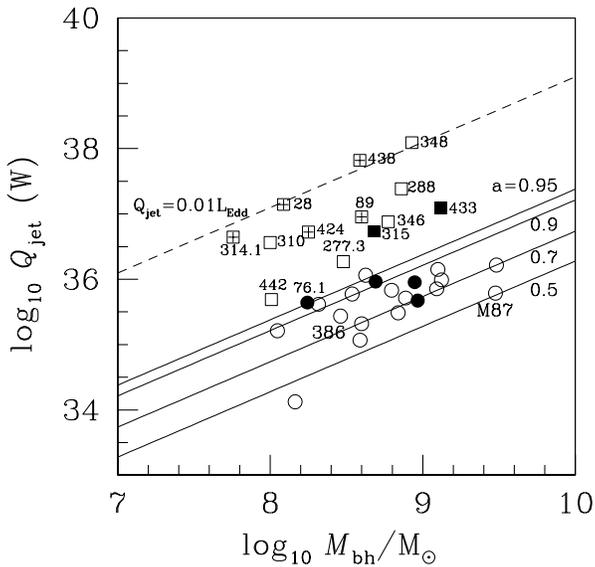,width=9.5cm,height=9.5cm}}
\caption{The black hole mass $M_{\rm bh}$ versus jet power $Q_{\rm
jet}$(assuming $f=1$). The solid lines represent the maximal jet
power extracted from the spinning black holes as functions of
$M_{\rm bh}$ for different labelled values of spin $a$. The
sources above the uppermost solid line (for $a=0.95$) are plotted
as squares, noted in Table 1, and called high-jet-power sources
throughout this paper. The filled symbols represent the sources
without detected central compact cores (see Table 1). The symbols with a
cross represent the sources with only upper limits on optical core fluxes.
The 3CR source names of the high-jet-power sources are labelled in
the figure, together with the objects `M87' and `3C386' discussed
in the text. } \label{mqj}
\end{figure}

\begin{figure}
\centerline{\psfig{figure=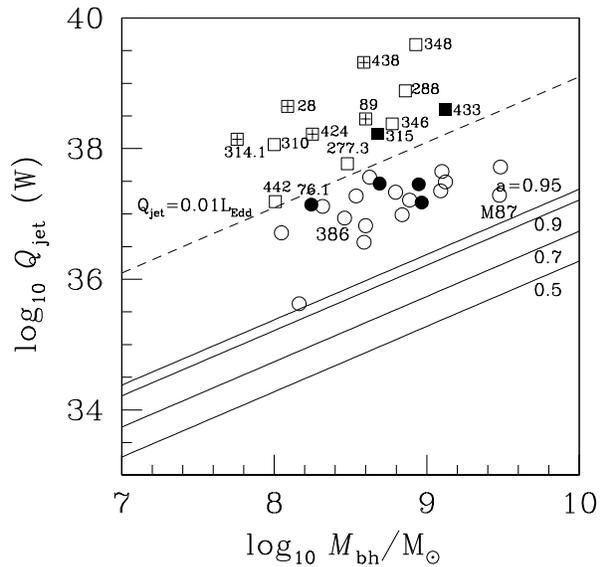,width=9.5cm,height=9.5cm}}
\caption{The same as Fig. \ref{mqj}, but with $f=10$ adopted in
the estimate of $Q_{\rm jet}$. } \label{mqj10}
\end{figure}

\begin{figure}
\centerline{\psfig{figure=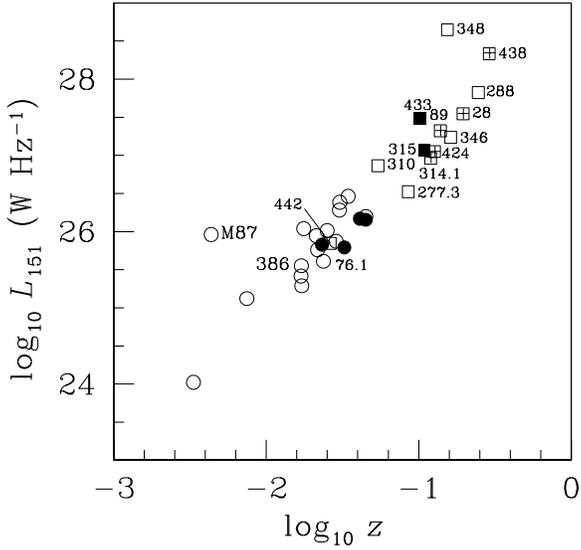,width=9.5cm,height=9.5cm}}
\caption{The total radio  luminosity at 151 MHz versus redshift. }
\label{zl151}
\end{figure}

\begin{figure}
\centerline{\psfig{figure=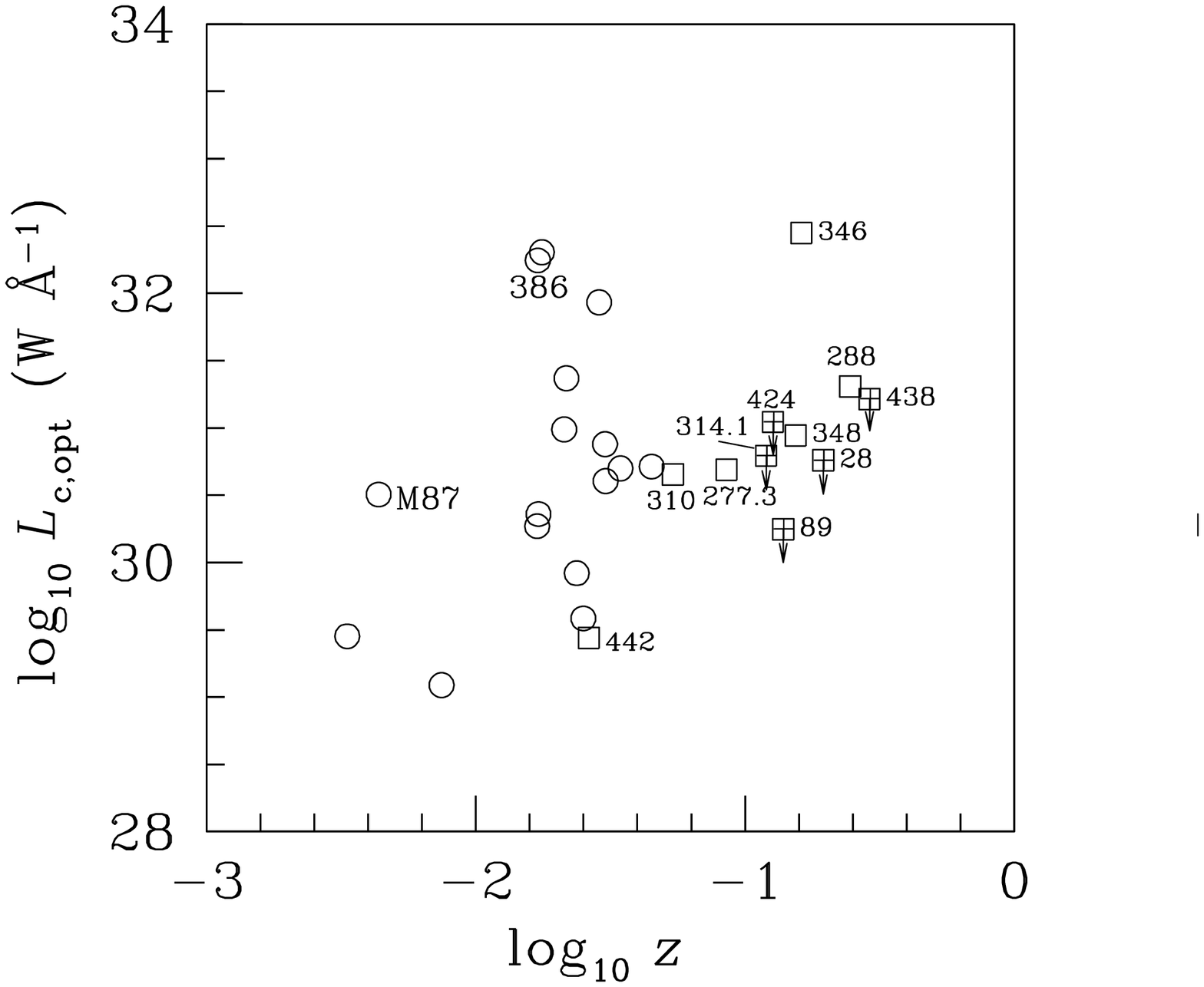,width=9.5cm,height=9.5cm}}
\caption{The optical core luminosity versus redshift. }
\label{zlcopt}
\end{figure}

\subsection{Optical core properties as a function of redshift}

The high-jet-power sources will also have significantly higher
redshifts than the other sources because jet power is estimated
from radio luminosity and radio luminosity correlates strongly
with redshift in flux-limited samples like 3CR (see Fig.
\ref{zl151}). The median redshift of the high-jet-power sources is
0.127, compared to 0.0235 for the low-jet-power sources, so the
HST linear size limits on any optical core emission are $\sim 5$
times larger for the high-jet-power sources (see Fig.
\ref{zlcopt}). Typically, therefore in low-jet-power sources, any
compact optical cores have sizes $<0.033$ kpc, whereas the
corresponding limit for high-jet-power sources is $<0.15$ kpc.

Note, also, that the objects with resolved optical cores (and
hence only limits on their optical core emission in Table 1; See
Sec. 2) are concentrated amongst the high-jet-power sources, as is
3C315 (see Sec. 2), the object with an elongated optical structure
and no optical core. It seems like a large fraction of the
high-jet-power, high-redshift sources have weak or absent cores,
and evidence for optical emission extended on $\sim$0.15 kpc
scales. It is not clear whether the extended optical cores of
these objects is associated with an extended synchrotron
component, or with host galaxy light. However, regardless, we can
conclude that the true optical cores of a large-fraction of the
high-jet-power sources may well be being obscured.

\subsection{Correlation between radio core luminosity and total
radio luminosity}

In Fig. \ref{l151l5}, we plot the relation between VLA radio core
luminosity at 5 GHz and the total radio luminosity at 151 MHz. We
find that a significant correlation between $L_{\rm c,5G}$ and
$L_{\rm 151}$, except two sources 3C28 and 3C314.1 have relatively
lower core luminosity. The generalized Kendall's $\tau$ test
(ASURV, Feigelson \& Nelson 1985) shows that the correlation is
significant at the 99.98\% level for the whole sample (the
significant level becomes $>$99.99\% for the sample subtracting
two sources 3C28 and 3C314.1). The jet power estimated from the
low-frequency radio luminosity represents the power of the jet
averaged over a long period of time ($>10^7$ years), while the VLA
core emission (typically corresponding to sub-arcsecond angular
size) reflects the emission from the jet near the central engine
which can obviously vary at a much shorter timescale ($\sim
10^{3-4}$ years). The correlation between $L_{\rm c,5G}$ and
$L_{\rm 151}$ indicates the jet power estimated from low-frequency
radio luminosity is a reliable measure of the current jet power in
most if not all the cases (e.g. 3C28 and 314.1). The key point
here is that with these possible exceptions (which could, for
example, be temporarily turned-off jets) the high-jet-power
sources probably have high jet powers throughout their life times,
and the physical model for how they are powered must account for
this.

\begin{figure}
\centerline{\psfig{figure=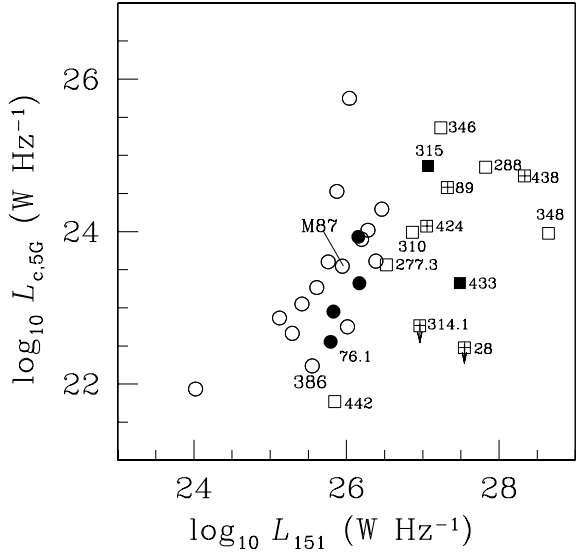,width=9.5cm,height=9.5cm}}
\caption{The radio core luminosity at 5~GHz versus total radio
luminosity at 151~MHz for the sources in the sample. }
\label{l151l5}
\end{figure}

\subsection{Optical core luminosity versus black hole mass}

We plot the optical core luminosity  $L_{\rm c,opt}$ versus black
hole mass $M_{\rm bh}$ in Fig. \ref{mlcopt}. The observed optical
core emission may be a combination of the synchrotron emission
from the jet and the emission from the disc, if the emission from
the disc has not been obscured.  The optical core luminosity
therefore sets an upper limit on the emission from the disc for
unobscured cases.  We convert the optical core luminosity to a
bolometric luminosity by using \citep{k00} \be L_{\rm bol}\simeq
9\lambda L_{\rm \lambda,opt}, \label{lbolopt}\ee which is valid
for normal bright AGNs. We plot the line corresponding to $L_{\rm
bol}=0.006L_{\rm Edd}$
(\citet{gc01}  proposed a diagonal line $L_{\rm ion}\sim 6\times
10^{-3} L_{\rm Edd}$ for the division of FR I and FR II galaxies).
It is found that all sources are, as expected, well below this
line, which indicates that most of these sources are accreting at
very low accretion rates, assuming of course that the disc
emission has not been obscured.

We note from the location of 3C386 in this plot, that some
optically unobscured objects with standard accretion discs, and
moderately high accretion rates, exist in the FR I population. We
have argued in Sect. 5.2 that, if we could account properly for
obscuration, a significant fraction, perhaps all, of the
high-jet-power sources might well lie in a similar region of this
plot. However, the strong correlation between the radio and
optical core luminosities for the low-jet-power objects, and the
restrictive limits the unresolved optical core places on any
obscuring material (e.g. $< 6$pc for M87), together imply that at
least some of the 3CR FR Is have genuinely low accretion rates, and
are unobscured.

\begin{figure}
\centerline{\psfig{figure=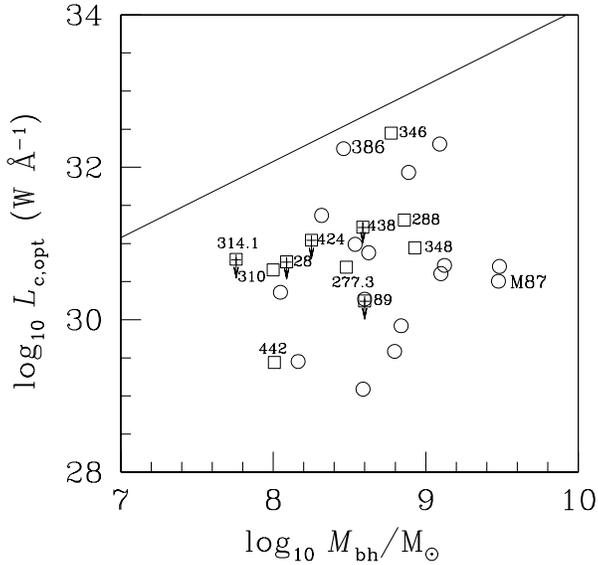,width=9.5cm,height=9.5cm}}
\caption{The black hole mass $M_{\rm bh}$ versus optical core
luminosity $L_{\rm c,opt}$. The line represents $L_{\rm
bol}=0.006~L_{\rm Edd}$. } \label{mlcopt}
\end{figure}

\subsection{Jet power versus optical core
luminosity}

The relation between the jet power $Q_{\rm jet}$ and optical core
luminosity $L_{\rm c,opt}$ is plotted in Fig. \ref{qjlcopt}. Usually, the
power in the jets of even radio-loud AGNs is only a fraction of their
bolometric luminosity (e.g., Rawlings \& Saunders 1991; Willott et al. 1999).
We adopt a parameter $\eta_{\rm jet}$ to relate the jet power
$Q_{\rm jet}$ to its bolometric luminosity $L_{\rm bol}$ \be
Q_{\rm jet}=\eta_{\rm jet} L_{\rm bol}\label{lboljet}.\ee Using
Eqs. (\ref{lbolopt}) and (\ref{lboljet}), we can calculate the
relation between the optical core luminosity and the jet power for
a given efficiency factor $\eta_{\rm jet}$. The lines for
different values of $\eta_{\rm jet}$ are plotted in Fig. \ref{qjlcopt}.

Note, first, the location of 3C386, the one object in our sample
where we definitely have an unobscured view of the nucleus. This
lies in a similar region of the plot to FR II radio sources (e.g.
Willott et al. 1999). The location of the high-jet-power sources
at high values of $\eta_{\rm jet} \sim 10$ is explicable if, as we
have postulated in Sec. 5.2, they have obscured nuclei. The
low-jet-power sources, which arguably are not obscured \citep{c99}
seem to lie reasonably close to $\eta_{\rm jet} \sim 1$, which may
be indicative of some difference between these objects and the
population of radio sources with more powerful jets.

\begin{figure}
\centerline{\psfig{figure=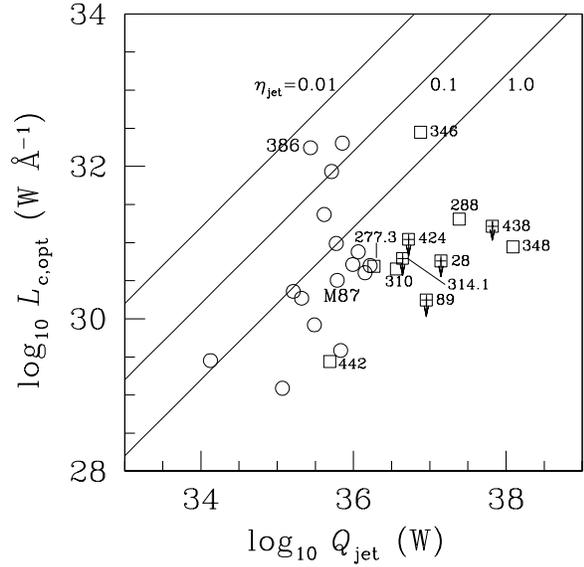,width=9.5cm,height=9.5cm}}
\caption{The jet power, calculated assuming $f=1$, versus the
optical core luminosity. The lines correspond to $\eta_{\rm
jet}=$0.01, 0.1 and 1, respectively where $\eta_{\rm jet}$ is
defined by Equation \ref{lboljet}. FR II radio sources would lie in the region
defined by the three parallel solid lines (e.g. Willott et al.
1999) } \label{qjlcopt}
\end{figure}

\subsection{Jet power versus line luminosity}

In Fig. \ref{qjll} we plot the relation between jet power and the
narrow-line luminosity. The generalized Kendall's $\tau$ test
(ASURV, Feigelson \& Nelson 1985) shows that the correlation is
significant at the 99.99\% level for the whole sample. The linear
regression by parametric EM Algorithm (ASURV) gives \be \log_{10}
L_{\rm line}=1.00(\pm0.16)\log_{10}Q_{\rm jet}-2.02(\pm5.69), \ee
for the whole sample but 3C348 is excluded (the dotted line in
Fig. \ref{qjll}), and \be \log_{10} L_{\rm
line}=1.03(\pm0.32)\log_{10}Q_{\rm jet}-3.22(\pm11.47), \ee for
the low-jet-power sources (the solid line in Fig. \ref{qjll}),
respectively.  It is found  that the high-jet-power FR I sources,
except 3C348, lie along the same correlation defined by the
low-jet-power FR I sources, which may imply that the narrow
emission-line formation mechanisms are similar for both these two
classes of the sources.

\begin{figure}
\centerline{\psfig{figure=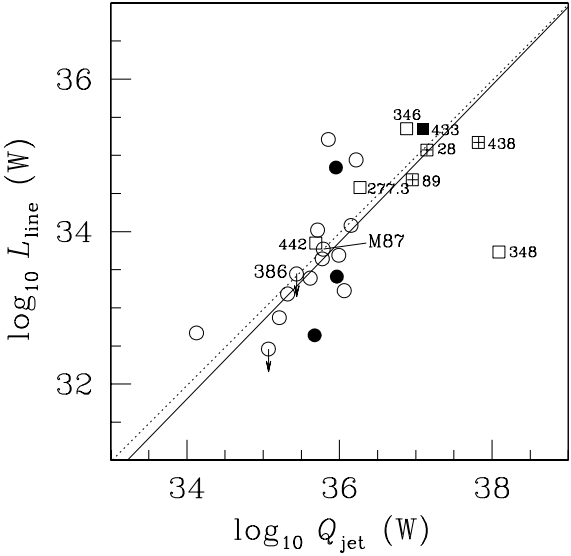,width=9.5cm,height=9.5cm}}
\caption{The relation between jet power  and narrow-line
luminosity. The dotted line represents the linear regression for
the whole sample but excluding 3C348,  while the solid line
represents the linear regression for the low-jet-power sources.}
\label{qjll}
\end{figure}

\subsection{Black hole mass versus line
luminosity $L_{\rm line}$}

In Fig. \ref{mll} we compare the different behaviours of the
high-jet-power FRIs and the low-jet-power FRIs in the narrow-line
luminosity versus black hole mass plane. The mean value of the
ratio of the  narrow-line luminosity to black hole mass is
$<\log_{10} L_{\rm line}/M_{\rm bh} (~\rm W~{\rm
M_\odot}^{-1})>=26.10\pm0.22$ for the high-jet-power sources,
while it is $<\log_{10} L_{\rm line}/M_{\rm bh}(~\rm
W~M_\odot^{-1})>=24.86\pm0.16$ for the low-jet-power sources.  The
separation of the two classes is clear, and is the direct result
of the strong correlation between radio and narrow-line
luminosities (Sec 5.6). There is no significant separation in
terms of black hole masses: the mean values of the black hole
masses are $<\log_{10} M_{\rm bh}/{\rm M_{\odot}}>=8.47\pm0.11$
and $<\log_{10} M_{\rm bh}/{\rm M_{\odot}}>=8.71\pm0.08$ for the
high-jet-power and low-jet power sources respectively. A natural
explanation is that the high emission line luminosity of these
high-jet-power sources is mainly contributed by power re-processed
from obscured optical nuclei. We therefore reach the same
conclusion as Section 5.2 by independent means: high-jet-power FRI
sources have relatively bright (i.e. as bright as 3C386), but
obscured nuclei.

\subsection{Size of the torus}

{The source 3C386 with an unobscured bright nucleus deviates
significantly from the correlation between radio and optical core
luminosities (see Fig. 5 in \citet{c99}). Most high-jet-power FRI
sources in our sample have higher line luminosities than 3C386
(see Fig. \ref{mll}), and one can infer that these high-jet-power
FRI sources have similar or brighter intrinsic optical nuclei than
3C386. The radio core luminosities of these high-jet-power FRI
sources are in the range of $\sim 10^{29}-10^{32}$ W~Hz$^{-1}$. If
the intrinsic bright nuclei were unobscured in these sources, as
is the case for 3C386, their locations in $L_{\rm r}-L_{\rm o}$
plane (Fig. 5 of \citet{c99}) would have definitely broken down
the linear correlation derived by \citet{c99}. This suggests that
only the synchrotron radio and optical emission from the jets are
detected in these high-jet-power FRI sources.

In the receding torus model, the inner radius of the torus is
roughly at the dust evaporation radius $r(T_{\rm evap})\simeq
0.06(L_{\rm bol}/10^{38}{\rm W})^{1/2}$ pc \citep{nl93,h96}. For
NGC4151, it predicts an evaporation radius $r(T_{\rm evap})\simeq
0.015$ pc ($L_{\rm bol}\simeq 6.5\times 10^{36}$ W, Kaspi et al.,
2000), which is roughly consistent with the inner radius of the
dust torus $\sim 0.04$ pc measured from a delay of the
near-infrared light curve to the optical light curve \citep{m03}.
The evaporation radius of the dust torus in 3C386 is estimated to
be $\sim 0.02$ pc ($L_{\rm bol}\sim 10^{37}$ W). The relative
thickness of the torus $s/r\sim 2-3$ is derived from the fraction
of quasars in the 3CR sample \citep{h96}. We can estimate the
total thickness $2s$ of the dust tori in these high-jet-power FRI
sources with obscured bright nuclei to be $\sim 0.05$ pc, because
the intrinsic nuclei in these sources are believed to be as bright
or brighter than  3C386. If the jet is inclined at a large angle,
for example, $\sim 45^\circ$, with respect to the line of sight,
the emission from the accretion disc should be obscured by the
torus while the synchrotron radio and optical emission from the
jet outside the torus can still be detected by VLBI at pc scales
or by HST at $\sim 0.01-0.1$ kpc scales. The emission at high
energy bands (e.g., hard X-ray bands) from the nucleus may
penetrate the obscuring material in the torus, and then  the
intrinsic properties of the nucleus can be explored directly from
its emission at high energy bands. However, this is applicable
only if the resolution is so high (e.g., $\sim 0.05$ pc) that the
emission from the jet outside the torus can be subtracted from the
emission of the nucleus. }

\begin{figure}
\centerline{\psfig{figure=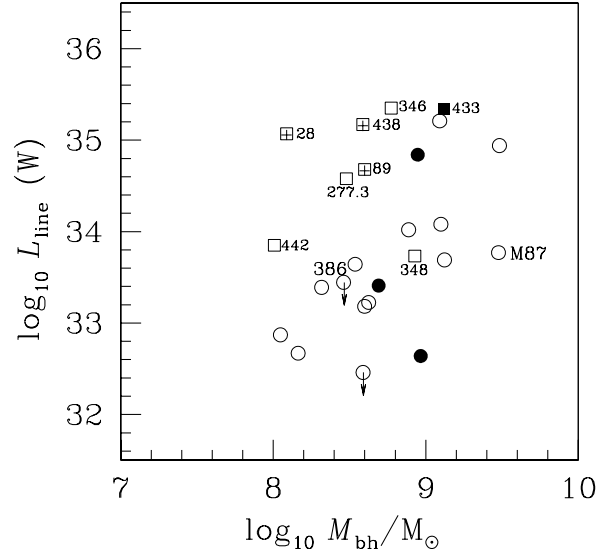,width=9.5cm,height=9.5cm}}
\caption{The black hole mass versus the narrow-line luminosity
plane.  } \label{mll}
\end{figure}

\section{Summary}

If all 3CR FRI radio sources were powered by central engines
operating in an ADAF mode, then the Blandford-Znajek mechanism
would have insufficient power to generate the high radio
luminosities of the most powerful FRI sources. Although it is
possible to postulate models in which an ADIOS mode is maintained
in such sources, this would still imply that, in some cases, a
significant fraction of the Eddington luminosity of a black hole
can be channelled into jets, with negligible emission at other
wavebands.

A less contrived solution to the existence of `high-jet-power' FRI
sources is to postulate that, in these FRI sources at least, there
are standard accretion discs. With the notable exception of 3C386,
these accretion discs seem typically to be hidden (see e.g.\
Grimes et al.\ 2003 for discussion of why this might be the case).
The large fraction of the high-jet-power FRI sources that lack any
unresolved HST optical core is consistent with this idea, as in
these cases the true nucleus is presumably hidden, and any optical
synchrotron is coming from scales larger than the obscuring
material. Also consistent with this idea is the strong correlation
between narrow-line and radio luminosities which means that, at
fixed black hole mass, high-jet-power FRI sources have
intrinsically brighter optical cores than low-jet-power FRI
sources, but these cores must be hidden

If the efficiency factor $f$ (see Equation \ref{qjetrad}) is nearer to $\sim
10$ than $\sim 1$ (as suggested by Blundell \& Rawlings 2000) then
these same arguments can probably be extended to include the
majority of 3CR FRI radio galaxies. Perhaps then, essentially all
of the 3CR FRIs have standard accretion discs. The case of M87
provides the best example of an object with a very low accretion
rate, and no torus bigger than $\sim 6$ pc. However, these two
factors alone are not sufficient to prove that M87 is accreting in
a regime described by an ADAF or ADIOS model. Indeed the location
of M87 in Fig. 3, suggests that, if $f \sim 10$, then its jets may
also be too powerful for it to be plausibly associated with an
ADAF mode of accretion unless there are powering mechanisms which
dominate over the Blandford-Znajeck process.

We conclude that the idea that the physical cause of the division
between FRI and FR II sources is that they have different modes of
accretion (e.g. Ghisellini \& Celotti 2001) is not supported by
the radio and optical properties of 3CR FRI radio galaxies. This
is in line with arguments made on the basis of individual objects
such as the 3CR FRI 3C386 (Simpson et al. 1996) and the
`radio-quiet' FRI quasar E1821+643 \citep{br01}.

\section*{Acknowledgments}
We are grateful for useful discussions with Jenny Grimes and Chris
Simpson, and helpful comments  by the referee, Neal Jackson. We
thank Marco Chiaberge and Danilo Marchesini for drawing our
attention to the superimposed star on the nucleus of 3C386. SR
thanks PPARC for a Senior Research Fellowship. XC thanks the
support from the Chinese Academy of Sciences and the Department of
Astrophysics, Oxford for its hospitality during his stay there.
The work is supported in part by the NSFC (No. 10173016; 10325314;
1033020) and NKBRSF (No. G1999075403). This research has made use
of the NASA/IPAC Extragalactic Database (NED), which is operated
by the Jet Propulsion Laboratory, California Institute of
Technology, under contract with the National Aeronautic and Space
Administration.

\section*{Note added in proof}

After the final version of this paper was submitted, we heard that
the nucleus of 3C386 is superimposed by a foreground star (D.
Marchesini, private communication). However,  a broad H$\alpha$
line is obviously seen in the spectrum of this source (Simpson et
al.,  1996),   The equivalent width of this broad line
corresponding to the central compact core emission measured  by
Chiaberge et al. (1999) is $\sim 15\rm\AA$, which implies that the
optical continuum emission from the nuclei of 3C386 is not
obscured and still quite bright, though it may not be so bright as
observed due to the contamination by the superimposed star. The
broad line and continuum emission could be variable as is common
in low radio luminosity broad-line radio galaxies (BLRGs) like
3C390.3. This may be verified by further optical spectroscopic
observations on 3C386.

\end{document}